\documentclass[reprint,amsmath,amssymb,aps]{revtex4-2} 

\usepackage{graphicx}
\usepackage{siunitx} 

\begin{document}

\title{The Dynamic Doppler Spectrum Induced by Nonlinear Sensor Motion: \\Relativistic Kinematics and 4D Frenet-Serret Spacetime Geometry}%

\author{Bryce M. Barclay}
\email[]{Bryce.Barclay@asu.edu}
\author{Alex Mahalov}
\email[]{Mahalov@asu.edu}
\affiliation{School of Mathematical and Statistical Sciences, Arizona State University}

\date{\today}

\begin{abstract}
Fundamental to the analysis of nonlinear relativistic motion is the precise characterization of the induced dynamic Doppler effects. 
In this work, we analyze the electromagnetic signals observed by non-inertial receivers using two frameworks to describe the relativistic motion.
We first consider observer paths described by higher-order kinematic 4-vectors: relativistic acceleration and jolt.
The dynamic Doppler effects of relativistic acceleration and jolt are exponential spectral broadening and exponential amplitude growth or decay. 
We derive compact expressions for the spectrum transformation resulting from relativistic acceleration and jolt. 
The jolt induces nonlinear skewed chirps in observed signals.
Next we consider observer paths described by the 4D Frenet-Serret frame and the curvature and torsion of the observer path. 
We obtain descriptions of the amplitude and phase fluctuations of the signal in terms of the geometric parameters of curvature and torsion. 
Concise, interpretable descriptions of non-inertial dynamic Doppler effects provide a useful diagnostic and predictive tool for engineering applications including radar, sensing, and communications systems. 
\end{abstract}

\maketitle

\section{Introduction\label{Introduction}}

The Doppler effect is a foundational element of many electromagnetic applications.
In tracking radar, changes in signal frequencies are used to estimate the motion of targets \cite{abratkiewicz2023target}.
In laser cooling, the Doppler effect is used to decrease the kinetic energy of particles \cite{kwolek2022continuous}. 
Recently, exploitation of Doppler effects was proposed to amplify ultrashort laser pulses \cite{de2025doppler,de2026scattering}.
The Doppler effect also forms a practical constraint on the design of wireless communications systems. 
With the growth of high-speed, highly-maneuverable objects in the modern aerial environment, a thorough assessment of complex Doppler effects is critical to the operation of electromagnetic infrastructure and the development of further applications. 
Continuous changes in velocity create dynamic Doppler effects which appear as chirps to observers.

Nonlinear sensor motions induce new physical effects that cannot be captured by the local constant velocity approximation alone. 
To provide a general correspondence between the dynamics of moving objects and the chirps produced by them, we carefully derive the electromagnetic signals observed by dynamic non-inertial objects. 
We utilize a geometric approach based on the Frenet-Serret frame of the observer which can be naturally described in 4D spacetime \cite{hari2024rotating,barclay2025spectral}. 

The Doppler shift was also shown to be fundamental to the Sagnac effect \cite{howell2022doppler}.
Variable-frequency signals are important in many areas of physics such as pulsar detection and dispersion measure searches \cite{jenet2000detection}.
As gravitational-wave detectors become more advanced, methods are being developed for signals with faint acceleration features \cite{chamberlain2019frequency}.
In many applications, long-time coherent integration can be implemented to increase the signal-to-noise ratio (SNR) \cite{huang2019long}, but this comes at the expense of introducing dynamic Doppler effects. 
Assessment of Doppler effects is increasingly critical to the operation of wireless communications networks, such as LEO satellite networks \cite{barclay2022sensor}.

Although relativistic considerations can be neglected in many engineering applications of the Doppler effect, relativity has played an important role in the implementation of many systems including the Global Positioning System \cite{ashby2003relativity,krieger2013relativistic}. 
Further, the physical theory of electromagnetic waves in non-inertial frames is well established in the literature \cite{neutze1998detecting,hauck2003electromagnetic,van1984relativity}, and it presents interesting physical phenomena which may be exploited to develop new technologies. 
In the development of sophisticated space-time signal processing methods which utilize the Doppler effect, the 4D relativistic framework is advantageous due to the Lorentz covariance of Maxwell's equations. 

In previous work \cite{barclay2024physics,barclay2024doppler}, we analyzed the spectra of signals received along nonlinear trajectories in 3D space. 
Here, we present a detailed analysis of the dynamic Doppler effects of nonlinear relativistic motion in 4D spacetime, i.e., the effects of relativistic acceleration, jolt, and movement in all 3 spatial dimensions on observed electromagnetic signals.
Higher-order kinematic vectors such as jolt are relevant for the study of mechanical shocks and for the control of acceleration (time-varying acceleration) in engineering applications. 
To this end, we fix a stationary signal transmitter with the inertial coordinate system $(x^\mu)$ and describe the kinematics of a non-inertial signal receiver in $(x^\mu)$ coordinates.
In terms of the receiver's world line $z(c\tau) = (ct_s(c\tau),x_s(ct(c\tau)))$, where $c$ is the speed of light, $x_s(ct)$ is the receiver trajectory in 3D space, and $\tau$ is the proper time of the receiver, the relativistic velocity, acceleration, and jolt are \cite{russo2009relativistic,pons2019observers}
\begin{align}
    u &= \frac{\partial z^\mu}{\partial c\tau}\frac{\partial}{\partial x^\mu}\\
    a &= \frac{\partial u^\mu}{\partial c\tau}\frac{\partial}{\partial x^\mu}\\
    \sigma &= j - a_\mu a^\mu u, \quad \text{ where } \quad j = \frac{\partial a^\mu}{\partial c\tau}\frac{\partial}{\partial x^\mu}. 
\end{align}
The electromagnetic tensor is the alternating covariant 2-tensor field:
\begin{align}
    F_{\mu\nu} = 
    \begin{pmatrix}
        0     & -E_x/c & -E_y/c & -E_z/c \\
        E_x/c & 0      & B_z    & -B_y   \\
        E_y/c & -B_z   & 0      & B_x    \\
        E_z/c & B_y    & -B_x   & 0  
    \end{pmatrix}.
\end{align}
To transform the electromagnetic field tensor to the perspective of the receiver, construct the frame $(e_{\mu'})$ of the receiver satisfying 
\begin{align}\label{Fermi_Walker_frame}
    e_{\mu'}\cdot e_{\nu'} = \eta_{\mu'\nu'},
    \qquad
    e_{0'} = u,\\
    \frac{d(e_{\lambda'})^\mu}{dc\tau} = (u^\mu a^\nu - u^\nu a^\mu)(e_{\lambda'})_\nu,\\
    e_{\mu'}(c\tau) = {\Lambda^\nu}_{\mu'}(c\tau)e_\nu,
\end{align}
where $(e_\mu)$ is the frame of the transmitter. 

The stationary transmitter with the coordinates $(x^\mu)$ that sends a continuous-wave electromagnetic communications signal given by 
\begin{align}
    E_{y} &= E_0\exp(i2\pi f_0(ct - \hat{k}_\ell x^\ell)/c) \label{transmitted_signal_E}\\
    B_{z} &= \frac{E_0}{c}\exp(i2\pi f_0(ct - \hat{k}_\ell x^\ell)/c),\label{transmitted_signal_B}
\end{align}
where the index $\ell$ indicates the spatial dimensions $1,2,3$ only, i.e., $\hat{k}_\ell x^\ell = \hat{k}_1 x^1 + \hat{k}_2 x^2 + \hat{k}_3 x^3$ (otherwise, indices include the temporal dimension $0$). The unit vector $\hat{k}$ is the propagation direction and $f_0$ is the frequency of the wave. 
We consider the microwave regime with $f_0 = 1~\si{GHz}$ as a reference value.  

This work is organized as follows. In section \ref{sec_spectral_analysis_kinematic}, the signal spectrum observed by a receiver with constant proper jolt and with constant proper acceleration are analyzed. In section \ref{sec_spectral_analysis_4DFSTS}, the spectrum is analyzed for non-inertial receivers using the Frenet-Serret geometric framework. Finally, in section \ref{sec_conclusion}, we discuss our results and present concluding remarks.

\section{Spectral analysis with relativistic kinematic parameters}\label{sec_spectral_analysis_kinematic}

In this section, we consider a transmitted signal propagating in the direction $\hat{k} = (1,0,0)$ and a receiver in the $x^0$-$x^1$ plane with constant proper jolt. 
The world line of the receiver is given by 
\begin{align}
    z^0(c\tau)
    &= \int_0^{c\tau} \gamma(\cosh(\omega(c\tau')) + \beta\sinh(\omega(c\tau')))dc\tau'\label{rela_jolt_path0}\\
    z^1(c\tau)
    &= \int_0^{c\tau} \gamma (\sinh(\omega(c\tau')) + \beta\cosh(\omega(c\tau')))dc\tau'\nonumber\\ &+ z^1(0).\label{rela_jolt_path1}
\end{align}
Here, $\beta = v_0/c$ is the relative initial speed of the receiver in $(x^\mu)$ coordinates, $\gamma = 1/\sqrt{1-\beta^2}$, and $\omega(c\tau) = a_0c\tau + j_0(c\tau)^2/2$ where $j_0 = |\sigma|$ and $a_0 = |a(0)|$. 
For simplicity, we set $z^1(0)=0$. 

Along the trajectory $z(c\tau)$, the receiver observes the signal given by Eqs. (\ref{transmitted_signal_E})-(\ref{transmitted_signal_B}).
The received phase and wavenumber functions are given by 
\begin{align}
    \Phi_s(c\tau) &= \\    
    \frac{2\pi f_0}{c}&\left(\sqrt{\frac{1-\beta}{1+\beta}}\int_0^{c\tau} \exp(-(a_0c\tau' + \frac{1}{2}j_0(c\tau')^2))dc\tau'\right)\nonumber\\
    K_s(c\tau) &= \frac{d\Phi_s}{dc\tau}(c\tau) = \sqrt{\frac{1-\beta}{1+\beta}}k_0\exp(-(a_0c\tau + \frac{1}{2}j_0(c\tau)^2)).\label{received_wavenumber_ts_jolt}
\end{align}
The received amplitude function is 
\begin{align}\label{received_amp_jolt}
    A_s(c\tau) = \sqrt{\frac{1-\beta}{1+\beta}} \exp(-(a_0c\tau + \frac{1}{2}j_0(c\tau)^2))E_0. 
\end{align}
Setting $a_0 = j_0 = 0$, we recover the standard relativistic Doppler frequency and amplitude factor:
\begin{align}
    D = \sqrt{\frac{1-\beta}{1+\beta}}.
\end{align}

Because phase fluctuations occur on shorter timescales than amplitude fluctuations, we use the stationary phase method (described in the appendix) to approximate the spectrum of the signal:
\begin{align}
    \int A(x^0)\exp(ik_0 h(x^0))\, dx^0
    &\approx \\
    \sum_m A(x^0_m)\left(\frac{2\pi}{-ik_0 h''(x^0_m)}\right)^{1/2}&\exp(ik_0 h(x^0_m)),\nonumber
\end{align}
where $x^0_m$ are the roots of $h'(x^0)$. 
To obtain the spectrum of the received signal as a function of wavenumber $k$, we use $k_0 h(x^0) = \Phi_s(x^0) - kx^0$.
The stationary phase method gives 
\begin{align}\label{SPA_rela_jolt0}
    \int &\exp(ik_0 h(x^0))\, dx^0
    \approx\\
    \sqrt{\frac{-i\pi}{k}}&\left(\frac{2}{j_0\log\left(\frac{Dk_0\exp(a_0^2/2j_0)}{k}\right)}\right)^{1/4}\exp(ik_0 h(x^0_+)) + \nonumber\\
    \sqrt{\frac{i\pi}{k}}&\left(\frac{2}{j_0\log\left(\frac{Dk_0\exp(a_0^2/2j_0)}{k}\right)}\right)^{1/4}\exp(ik_0 h(x^0_-)),\nonumber
\end{align}
where the stationary points $x^0_\pm$ are 
\begin{align}
    x^0_\pm = -\frac{a_0}{j_0} \pm \sqrt{\frac{2\log\left(\frac{Dk_0\exp(a_0^2/2j_0)}{k}\right)}{j_0}}.
\end{align}
In general, the contributions to the spectral approximation (\ref{SPA_rela_jolt0}) from the two stationary points create an interference pattern in the spectrum. 

In the remainder of this section, we analyze the spectrum of the received signal over time intervals $[c\tau_i,c\tau_f]$ where $K_s(c\tau)$ is monotonic and only one stationary point contribution is present, i.e.,
\begin{align}\label{SPA_rela_jolt}
    \int &\mbox{rect}(x^0)\exp(ik_0 h(x^0))\, dx^0
    \approx\\
    \sqrt{\frac{-i\pi}{k}}&\left(\frac{2}{j_0\log\left(\frac{Dk_0\exp(a_0^2/2j_0)}{k}\right)}\right)^{1/4}\exp(ik_0 h(x^0_+)) \nonumber
\end{align}
where $\mbox{rect}(x^0) = 1$ for $c\tau_i\leq x^0 \leq c\tau_f$ and $0$ otherwise.

Using $A = A_s\mbox{rect}$ where $A_s$ is the received amplitude function given by Eq. (\ref{received_amp_jolt}), the amplitude spectrum of the received signal is 
\begin{align}\label{SPA_rela_jolt_amp}
    |S(k)| = \Bigg|\int &A(x^0)\exp(ik_0 h(x^0))\, dx^0\Bigg|
    \approx\\
    &E_0\left(\frac{2\pi^2k^2}{k_0^4j_0\log\left(\frac{Dk_0\exp(a_0^2/2j_0)}{k}\right)}\right)^{1/4}.\nonumber
\end{align}

Substituting the wavenumber function $k = K_s(c\tau)$ gives 
\begin{align}
    |S(K_s(c\tau))| &= \\ 
    E_0\left(\frac{2\pi^2D^2}{k_0^2j_0}\right)^{1/4}&\frac{\exp\left(-\frac{1}{2}(a_0c\tau + \frac{1}{2}j_0(c\tau)^2)\right)}{\left(\frac{j_0}{2}(c\tau)^2 + a_0c\tau + \frac{a_0^2}{2j_0}\right)^{1/4}}.\nonumber
\end{align}
The ratio of the amplitude spectrum at $c\tau=c\tau_f$ to $c\tau=0$ is
\begin{align}
    A_{j}(c\tau_f) = \frac{|S(K_s(c\tau_f))|}{|S(K_s(0))|} = 
    \frac{\exp\left(-\frac{a_0^2}{4j_0}\left(\eta^2-1\right)\right)}{\sqrt{\eta}},
\end{align}
where $\eta = \frac{j_0}{a_0}c\tau_f + 1$. 
The ratio of the signal wavenumber at $c\tau=c\tau_f$ to $c\tau=0$ is 
\begin{align} \label{jolt_wavenumber_factor}
    D_{j}
    &=
    \exp\left(-\frac{a_0^2}{2j_0}\left(\eta^2-1\right)\right).
\end{align}
Then the amplitude factor is 
\begin{align}\label{jolt_amplitude_factor}
    A_j(c\tau_f) = \sqrt{\frac{D_{j}}{\eta}}.
\end{align}

\begin{figure}
	\centering
    \includegraphics[width=85mm]{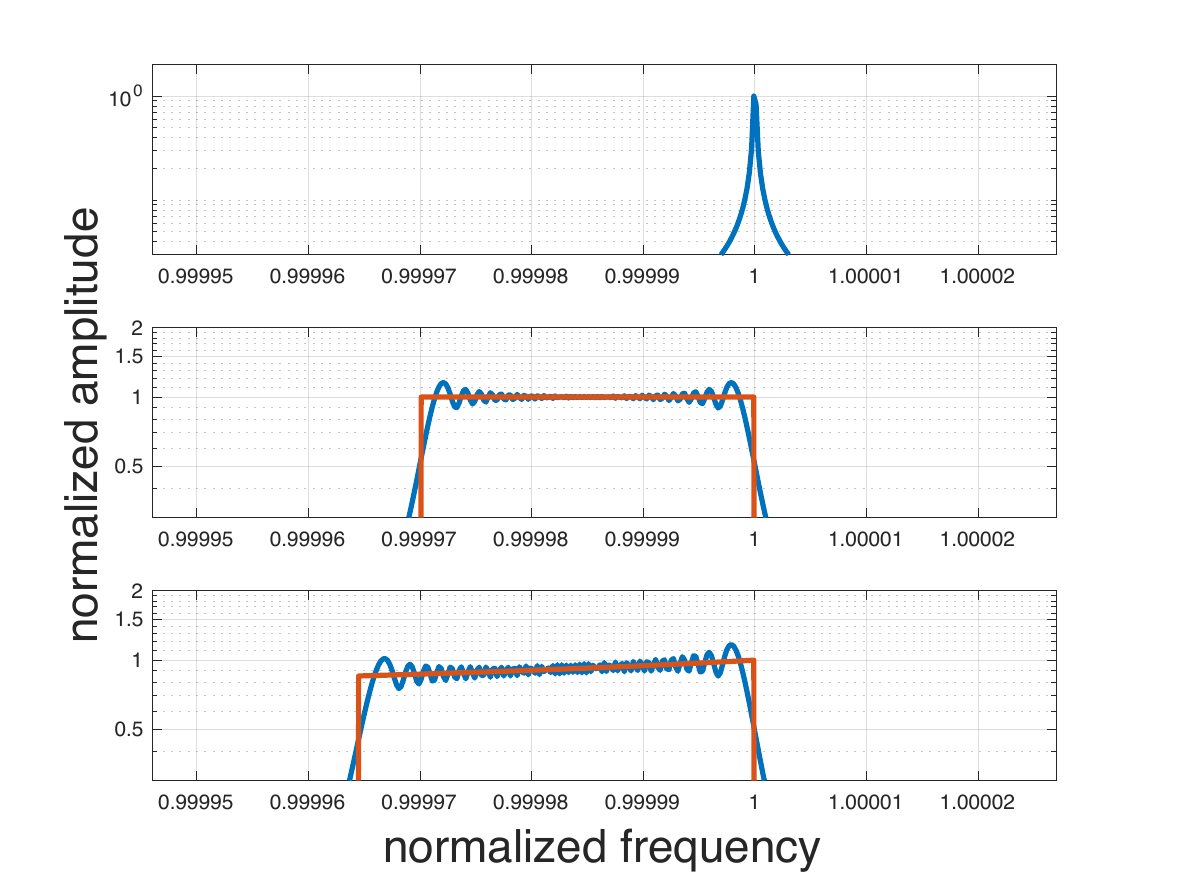}
	\caption{Amplitude spectra of the signal received along the spacetime path $z^\mu(c\tau)$ given by Eqs (\ref{rela_jolt_path0})-(\ref{rela_jolt_path1}) in blue. The frequency and amplitude are normalized by their respective values at $c\tau=0$. The spectrum approximation is obtained using the stationary phase method (\ref{SPA_rela_jolt_amp}). In the top panel, proper jolt $j_0$ and proper acceleration $a_0$ are zero (classical Doppler shift). In the middle panel (chirp spectrum), proper jolt $j_0$ is zero. In the bottom panel (skewed chirp spectrum), proper jolt $j_0$ is nonzero. In the bottom two panels, the stationary phase approximation is displayed in red. }
	\label{fig_relative_freq_relative_amp}
\end{figure}

\begin{figure}
	\centering
    \includegraphics[width=85mm]{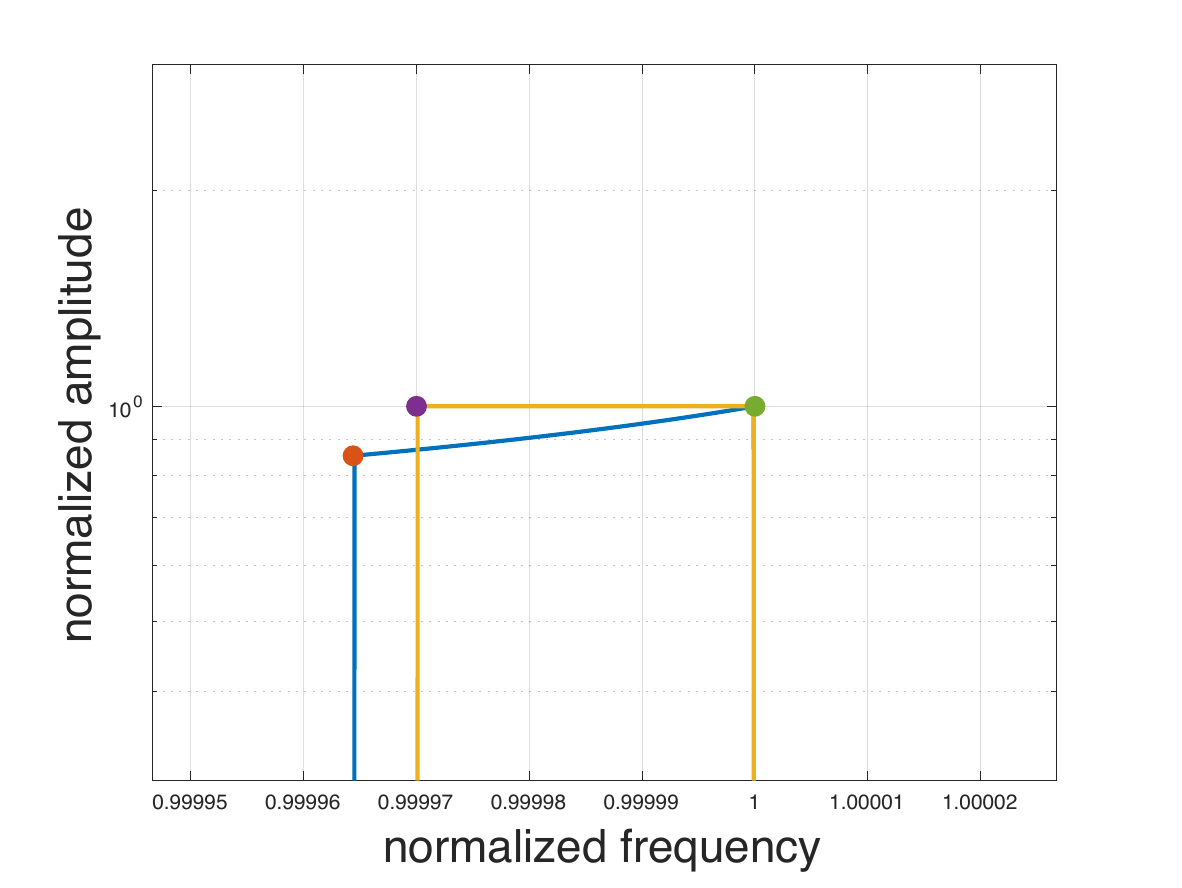} 
	\caption{Amplitude spectra of the signal received along the spacetime path $z^\mu(c\tau)$ given by Eqs (\ref{rela_jolt_path0})-(\ref{rela_jolt_path1}). 
    The frequency and amplitude are normalized by their respective values at $c\tau=0$. 
    The spectrum approximation is obtained using the stationary phase method (\ref{SPA_rela_jolt_amp}). 
    The yellow and blue curves correspond to constant proper acceleration and constant proper jolt, respectively. For constant proper acceleration, the amplitude spectrum is nearly constant. 
    The amplitude spectrum is strongly nonlinear for a receiver with constant proper jolt. 
    }
	\label{fig_relative_freq_relative_amp_2}
\end{figure}

When the proper jolt $j_0$ is zero, the world line of the receiver is given by Eqs. (\ref{rela_jolt_path0})-(\ref{rela_jolt_path1}) 
where $\omega(c\tau) = a_0c\tau$ and $a_0 = |a(0)|$.  
The received phase and wavenumber functions are given by 
\begin{align}
    \Phi_s(c\tau) &= \frac{-k_0 D}{a_0}\exp(-a_0c\tau) \\
    K_s(c\tau) &= \frac{d\Phi_s}{dc\tau}(c\tau) = k_0 D\exp(-a_0c\tau).
\end{align}
The ratio of the signal wavenumber at $c\tau=c\tau_f$ to $c\tau=0$ is 
\begin{align} \label{accel_waveumber_factor}
    D_{a} = \exp(-a_0c\tau_f).
\end{align}
The amplitude spectrum is given by
\begin{align}
    \Bigg|\int &\exp(ik_0 h(x^0))\, dx^0\Bigg|
    \approx
    \sqrt{\frac{2\pi}{a_0k}}
    =
    \sqrt{\frac{2\pi\exp(a_0c\tau)}{a_0k_0 D}}.
\end{align}
The ratio of the amplitude at $c\tau=c\tau_f$ and $c\tau=0$ is 
\begin{align} \label{accel_amplitude_factor}
    A_{a}(c\tau_f) = \exp(-a_0c\tau_f/2) = \sqrt{D_{a}}.
\end{align}
The amplitude spectra for constant proper jolt and constant proper acceleration are compared in Figs. \ref{fig_relative_freq_relative_amp} and \ref{fig_relative_freq_relative_amp_2}. 
In Fig. \ref{fig_relative_freq_relative_amp_2}, the green point can be mapped to the purple point via the transformation $(k,S) \mapsto (D_{a}k,A_{a}S)$ or mapped to the red point via the transformation $(k,S) \mapsto (D_{j}k,A_{j}S)$. 
The jolt of a receiver accelerating away from the signal emitter induces a nonlinear, skewed chirp where the amplitude decays as the frequency shifts down. For constant proper acceleration, the amplitude is nearly constant as a function of frequency. 

\section{Spectral analysis with the Frenet-Serret Frame in 4D Spacetime}\label{sec_spectral_analysis_4DFSTS}

The receiver path can be approximated by a series expansion in terms of the geometric parameters of curvature, torsion, and hyper-torsion of the 4D path using the Frenet-Serret frame:
\begin{align}\label{path_4DFSTS}
    z(c\tau) 
    &= 
    z(0) + z'(0)c\tau + \frac{z''(0)}{2}(c\tau)^2 + \frac{z'''(0)}{6}(c\tau)^3\nonumber\\
    &+ \frac{z^{(4)}(0)}{24}(c\tau)^4 + \dots\nonumber\\
    &=
    z(0) + \left(c\tau + \frac{\kappa_1^2}{6}(c\tau)^3 + \frac{3\kappa_1\kappa_1'}{24}(c\tau)^4\right)e_{0''}\nonumber\\ 
    &+ \left(\frac{\kappa_1}{2}(c\tau)^2 + \frac{\kappa_1'}{6}(c\tau)^3 + \frac{\kappa_1^3 + \kappa_1'' - \kappa_1\kappa_2^2}{24}(c\tau)^4\right)e_{1''}\nonumber\\ 
    &+ \left(\frac{\kappa_1\kappa_2}{6}(c\tau)^3 + \frac{(2\kappa_1'\kappa_2 + \kappa_1\kappa_2')}{24}(c\tau)^4\right)e_{2''}\nonumber\\ 
    &+ \left(\frac{\kappa_1\kappa_2\kappa_3}{24}(c\tau)^4\right)e_{3''} + \dots,
\end{align}
where derivatives are taken with respect to $c\tau$ and the parameters $\kappa_j$ and vectors $e_{k''}$ are evaluated at $c\tau=0$. 
The Frenet-Serret vectors are given in 4D spacetime by 
\begin{align} 
    \frac{de_{0''}}{dc\tau} &=  \kappa_1e_{1''}\label{FS1}\\
    \frac{de_{1''}}{dc\tau} &=  \kappa_1e_{0''} + \kappa_2 e_{2''}\label{FS2}\\
    \frac{de_{2''}}{dc\tau} &= -\kappa_2e_{1''} + \kappa_3 e_{3''}\label{FS3}\\
    \frac{de_{3''}}{dc\tau} &= -\kappa_3e_{2''},\label{FS4}
\end{align} 
where $e_{0''}$ corresponds to the 4-velocity and is tangent to the path, and the three spacelike vectors are called the normal, binormal, and trinormal frame vectors. The geometric parameters $\kappa_1$, $\kappa_2$, and $\kappa_3$ are the curvature, torsion, and hyper-torsion of the receiver trajectory. 

Truncating to third order gives the approximate path
\begin{align}
    z(c\tau) 
    \approx
    z(0) &+ \left(c\tau + \frac{\kappa_1^2}{6}(c\tau)^3\right)e_{0''}\nonumber\\ 
    &+ \left(\frac{\kappa_1}{2}(c\tau)^2 + \frac{\kappa_1'}{6}(c\tau)^3\right)e_{1''}\nonumber\\ 
    &+ \left(\frac{\kappa_1\kappa_2}{6}(c\tau)^3\right)e_{2''}.
\end{align}
For simplicity, we set $z(0) = 0$. 
The 3-velocity of the observer in the frame of the signal transmitter is given by 
\begin{align}
    v(c\tau) 
    &= c\beta(c\tau) 
    = c\frac{dx_s/dc\tau}{dct/dc\tau}.
\end{align}

The received phase function for the emitted plane wave is
\begin{align}
    \Phi_s(c\tau) 
    = k_0k_\mu z^\mu(c\tau)
    &= k_0k_\mu \zeta^{\nu''}(c\tau){e_{\nu''}}^\mu(c\tau)\nonumber\\
    &= k_0\zeta^{\nu''}(c\tau)\alpha_{\nu''}(c\tau)
\end{align}
where $k_\mu = (1,-\hat{k}_1,-\hat{k}_2,-\hat{k}_3)$ and $\zeta^{\nu''}(c\tau)$ are the coefficients of $z(c\tau)$ with respect to the $e_{\nu''}$ frame. 
Here we define $\alpha_{\nu''}(c\tau) = k_\mu {e_{\nu''}}^\mu(c\tau)$ and $w^\mu = (\kappa_1^2,\kappa_1',\kappa_1\kappa_2,0)$. 
Using the series expansion (\ref{path_4DFSTS}), the received phase is 
\begin{align}
    \Phi_s(c\tau) 
    &= k_0\alpha_{\nu''}(0)\zeta^{\nu''}(c\tau)
\end{align}
and the received wavenumber function is 
\begin{align}
    K_s(c\tau) 
    &= k_0\alpha_{\nu''}(0)\frac{d\zeta^{\nu''}}{dc\tau}(c\tau)\nonumber\\
    &= k_0\bigg(\alpha_\mu w^\mu \frac{(c\tau)^2}{2} + \alpha_1\kappa_1c\tau + \alpha_0\bigg).
\end{align}

Because the wavenumber function is not necessarily monotonic, we consider the monotonic and non-monotonic cases separately, beginning with the monotonic case.
For the third-order observer path expansion, the stationary phase method gives 
\begin{align}\label{SPA_4DFSTS}
    &\int \exp(ik_0 h(x^0))\, dx^0
    \approx\\
    &\mbox{\normalsize $\frac{\sqrt{-2\pi i}\exp(ik_0 h(x^0_+))}{(k_0(\alpha_1^2\kappa_1^2k_0 + 2(\alpha_\mu w^\mu) (k-\alpha_0k_0)))^{1/4}} +$} \nonumber\\
    &\mbox{\normalsize $\frac{\sqrt{2\pi i}\exp(ik_0 h(x^0_-))}{(k_0(\alpha_1^2\kappa_1^2k_0 + 2(\alpha_\mu w^\mu) (k-\alpha_0k_0)))^{1/4}}$}.\nonumber
\end{align}

\begin{figure}
	\centering
	\includegraphics[width=85mm]{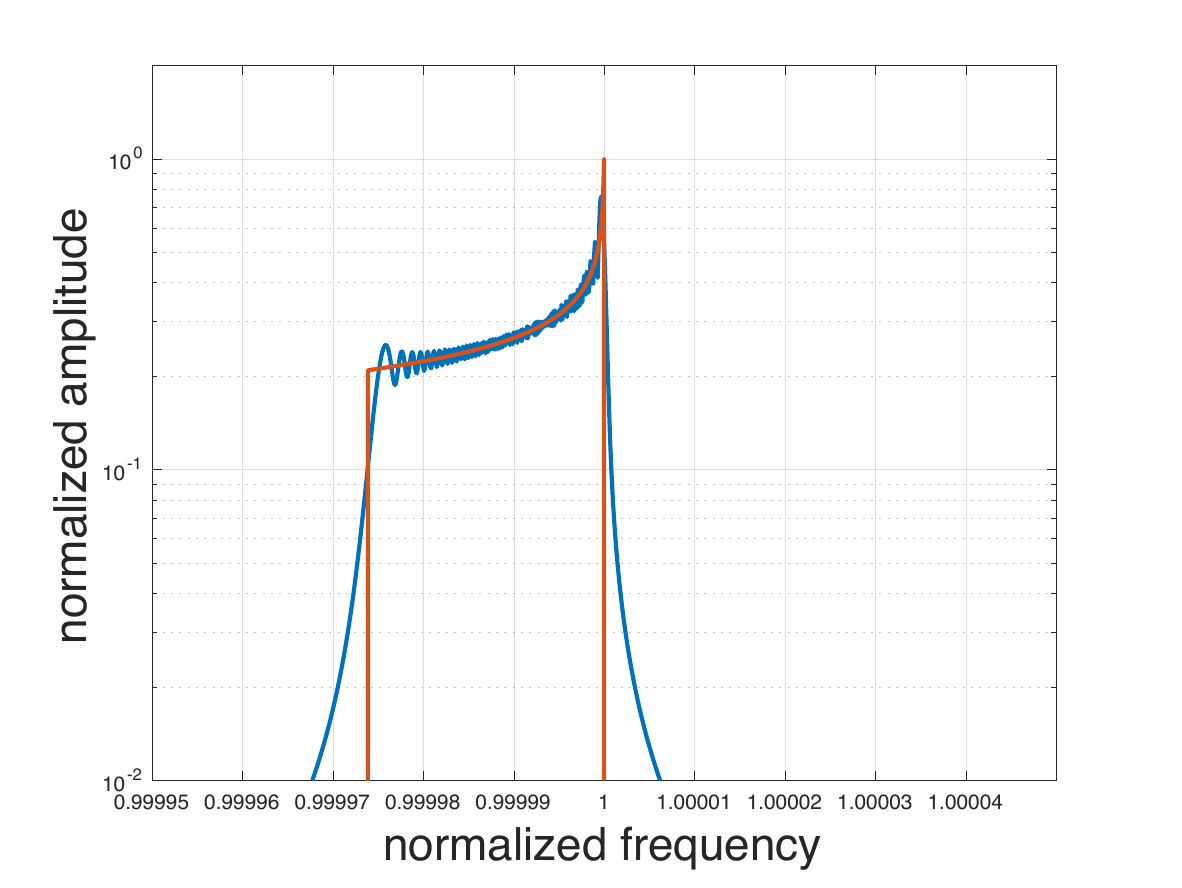}
	\caption{Comparison of the spectrum of the signal received along the spacetime path $z^\mu(c\tau)$ given by Eq. (\ref{path_4DFSTS}) (in blue) to the spectrum approximation (\ref{SPA_4DFSTS}) obtained from the stationary phase method (in red). The receiver path $z^\mu(c\tau)$ is described by the 4D Frenet-Serret expansion. }
	\label{fig_4DFSTS_spectrum}
\end{figure}

Further,
\begin{align}
    h_0 &= 
    \frac{h(x_+^0) + h(x_-^0)}{2} \\
    &= \frac{\alpha_1\kappa_1(\alpha_1^2\kappa_1^2k_0 + 3(\alpha_\mu w^\mu) (k-\alpha_0k_0))}{3k_0(\alpha_\mu w^\mu)^2}\nonumber\\
    h_1 &= 
    \frac{h(x_+^0) - h(x_-^0)}{2} \\
    &= \frac{(\alpha_1^2\kappa_1^2k_0 + 2(\alpha_\mu w^\mu) (k-\alpha_0k_0))^{3/2}}{3k_0^{3/2}(\alpha_\mu w^\mu)^2}.\nonumber
\end{align}
For a monotonic wavenumber function $K_s(c\tau)$, the spectrum approximation consists of one stationary point: 
\begin{align}\label{SPA_4DFSTS_mono}
    &\int \mbox{rect}(x^0)\exp(ik_0 h(x^0))\, dx^0
    \approx\\
    &\mbox{\normalsize $\frac{\sqrt{-2\pi i}\exp(ik_0 h(x^0_+))}{(k_0(\alpha_1^2\kappa_1^2k_0 + 2(\alpha_\mu w^\mu) (k-\alpha_0k_0)))^{1/4}}$}. \nonumber
\end{align}
The approximation given by Eq. (\ref{SPA_4DFSTS_mono}) is displayed in Fig. \ref{fig_4DFSTS_spectrum}. 
The result of acceleration along a nonlinear spatial trajectory $x_s(ct)$ is a nonlinear, skewed chirp with an amplitude decay as the frequency shifts down similar to the chirp induced by jolt discussed in section \ref{sec_spectral_analysis_kinematic}. 

The ratio of the amplitude at $c\tau = c\tau_f$ and $c\tau = 0$ is 
\begin{align}
    A_{FS}(c\tau_f) = 
    \frac{1}{\sqrt{\eta}},
\end{align}
where $\eta = \frac{\alpha_\mu w^\mu}{\alpha_1\kappa_1}c\tau_f + 1$.
The ratio of the signal wavenumber at $c\tau = c\tau_f$ and $c\tau = 0$ is 
\begin{align}
    D_{FS} = 
    \frac{(\alpha_1\kappa_1)^2}{2\alpha_0(\alpha_\mu w^\mu)}\left(\eta^2 - 1\right) + 1.
\end{align}

The wavenumber function is non-monotonic around the critical value $k_c$: 
\begin{align}
    k_c = k_0\bigg(\alpha_0 - \frac{(\alpha_1\kappa_1)^2}{2\alpha_\mu w^\mu}\bigg). 
\end{align}

\begin{figure}
	\centering
	\includegraphics[width=85mm]{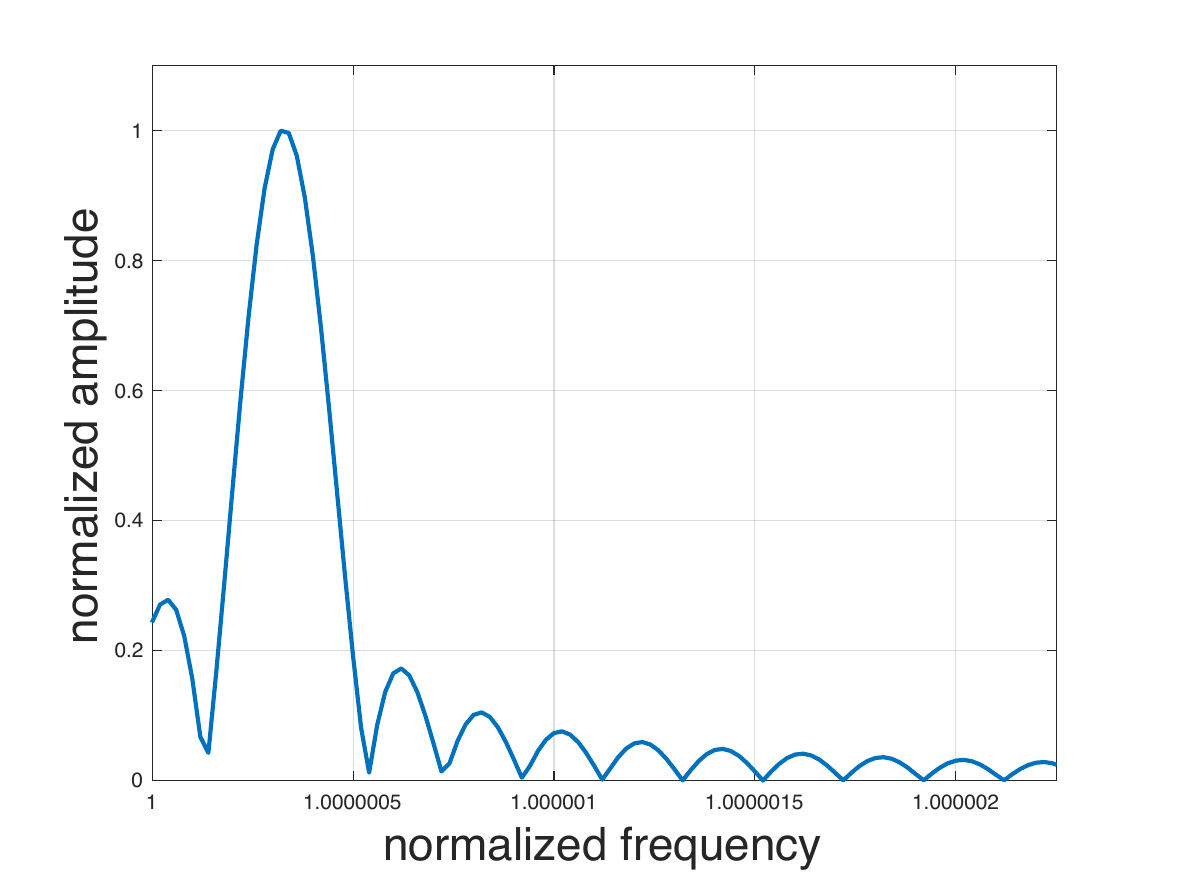} 
	\caption{The Airy approximation (\ref{Airy_spectrum_approx}) of the spectrum of the signal received along the 4D Frenet-Serret path $z^\mu(c\tau)$ given by Eq. (\ref{path_4DFSTS}). The received signal frequency is a non-monotonic function of time creating an interference pattern in the spectrum. }
	\label{fig_4DFSTS_spectrum_Airy}
\end{figure}

To provide a uniform approximation for the non-monotonic case, the Airy function approximation is used (see the appendix): 
\begin{align}\label{Airy_spectrum_approx}
    \frac{2\pi\exp(ik_0h_0)}{(\alpha_\mu w^\mu k_0/2)^{1/3}}\mbox{Ai}\left(-\left(\frac{3}{2}k_0h_1\right)^{2/3}\right).
\end{align}
The amplitude spectrum for this scenario is displayed in Fig. \ref{fig_4DFSTS_spectrum_Airy}. 
Because the frequency of the received signal is a non-monotonic function of time, different times corresponding to the same frequency interfere, creating amplitude fluctuations in the spectrum. 

\section{Conclusion}\label{sec_conclusion}

In this work, we presented a thorough analysis of the spectral artifacts of non-inertial observer motion. 
We derived interpretable expressions for the dynamic Doppler shift and spectral amplitude growth (or decay). 
As further Doppler-based technologies are established, precise quantitative descriptions of the nonlinear spectral features of emitted and observed signals will be necessary.
These spectral features can be utilized to develop novel physics-informed signal processing, AI, and machine learning algorithms for Doppler mitigation and/or exploitation strategies in next-generation communications, sensing, and radar applications.  
This work establishes building blocks from which general motions in 4D spacetime can be delineated. 
Further development includes the analysis of Bez\'ier curves in 4D spacetime and the creation of algorithms to detect dynamic Doppler effects induced by motion along 4D Bez\'ier curves. 

\begin{acknowledgments}
This material is based upon work supported by the Air Force Office of Scientific Research under award number FA9550-23-1-0177. 
\end{acknowledgments}

\bibliography{refs}

\appendix
\section{The method of stationary phase and the Airy framework}
The stationary-phase approximation of an oscillatory integral is given by
\begin{align}\label{stationary_phase_approximation}
	\int A(t)\exp(i\lambda h(t))\, dt
	&\approx\\
	\sum_m &A(t_m)\left(\frac{2\pi}{-i\lambda h''(t_m)}\right)^{1/2}\exp(i\lambda h(t_m)),\nonumber
\end{align}
where the stationary times $t_m$ are the roots of $h'(t)$. 

We now provide the stationary-phase approximation of an oscillatory integral on a finite time interval $[T_1,T_2]$:
\begin{align}\label{cubic_phase_integral}
	\int_{T_1}^{T_2} \exp(i\lambda h(t))\, dt,
\end{align}
with a general, cubic phase $h(t) = at^3 + bt^2 + ct$, where $a > 0$, $b$, and $c$ are any fixed constants. Integration over the finite interval $[T_1,T_2]$ is represented by setting $A(t) = 1$ for $t$ in $[T_1,T_2]$ and setting $A(t) = 0$ otherwise in (\ref{stationary_phase_approximation}). The approximation can be divided into three cases depending on the number and multiplicity of the stationary points contained in the interval of integration $[T_1,T_2]$. The stationary points of $h(t)$ are the roots of $h'(t)$, denoted by $t_\pm$. The first case is when $T_1 < t_- < t_+ < T_2$. The stationary-phase approximation is
\begin{align}
	\int_{T_1}^{T_2} \exp(i\lambda h(t))\, dt
	&\approx 
	\pi^{1/2}\frac{\exp(i(\lambda h(t_-) - \pi/4))}{[b^2 - 3ac]^{1/4}}\lambda^{-1/2}\nonumber\\
	&+ 
	\pi^{1/2}\frac{\exp(i(\lambda h(t_+) + \pi/4))}{[b^2 - 3ac]^{1/4}}\lambda^{-1/2},\nonumber
\end{align}
where the boundary terms of order $\lambda^{-1}$:
\begin{align}
    \frac{\exp(i(\lambda h(T_1) + \pi/2))}{3a(t_--T_1)(t_+-T_1)}\lambda^{-1}
	&+ 
	\frac{\exp(i(\lambda h(T_2) - \pi/2))}{3a(T_2-t_-)(T_2-t_+)}\lambda^{-1}\nonumber
\end{align}
are omitted. 
For the cubic phase $h(t)$, the local extrema $h(t_\pm)$ can be written explicitly:
\begin{align}
	h(t_\pm) 
	&= 
	\frac{2b^3 - 9abc \mp 2[b^2 - 3ac]^{3/2}}{27a^2} = h_0 \mp h_1.
\end{align}
Then the stationary phase approximation is
\begin{align}
	\int_{T_1}^{T_2} \exp(i\lambda h(t))\, dt
	&\approx
	\frac{2\pi^{1/2}}{(3a)^{1/3}}\frac{\cos(\lambda h_{1} - \pi/4)\exp(i\lambda h_0)}{\left(\frac{3}{2}h_1\right)^{1/6}}\lambda^{-1/2}.\nonumber
\end{align} 
See \cite{erdelyi1955asymptotic} and \cite{stein1993harmonic}, ch.~8 for more information on the stationary-phase method. 
The second case is when $T_1 < t_- = t_+ < T_2$. It is useful in this case to provide an approximation which is uniformly valid when $t_+ - t_-$ is near zero. To accomplish this, the integral (\ref{cubic_phase_integral}) is approximated by Airy functions using the transformation:  
\begin{align}
	h(t) = h_0 - \left(\frac{3}{2}h_{1}\right)^{2/3}s + \frac{1}{3}s^3,
\end{align}
where
\begin{align}
	\frac{dt}{ds} 
	&= 
	\frac{s^2 - \left(\frac{3}{2}h_{1}\right)^{2/3}}{3at^2 + 2bt + c}\\
	&=
	\sum_{m=0}^\infty p_m\left(s^2 - \left(\frac{3}{2}h_{1}\right)^{2/3}\right)^m + q_ms\left(s^2 - \left(\frac{3}{2}h_{1}\right)^{2/3}\right)^m,\nonumber
\end{align}
giving 
\begin{align}
	\int_{T_1}^{T_2} \exp(i\lambda h(t))\, dt
	&\approx
	\frac{2\pi\exp(i\lambda h_0)}{(3a\lambda)^{1/3}}\mbox{Ai}\left(-\left(\frac{3}{2}\lambda h_{1}\right)^{2/3}\right).
\end{align}

The third case is when only one root is in the interval, which for simplicity we will assume is $t_-$. The stationary-phase approximation in this case is
\begin{align}
	\int_{T_1}^{T_2} \exp(i\lambda h(t))\, dt
	&\approx
	\frac{\pi^{1/2}}{(3a)^{1/3}}\frac{\exp(i(\lambda (h_0+h_1) - \pi/4))}{\left(\frac{3}{2}h_1\right)^{1/6}}\lambda^{-1/2}.\nonumber
\end{align}

\end{document}